\documentstyle[12pt,fleqn,epsfig]{article}

\newlength{\dinwidth}
\newlength{\dinmargin}
\def\lapproxeq{\lower .7ex\hbox{$\;\stackrel{\textstyle<}{\sim}\;$}}
\def\gapproxeq{\lower .7ex\hbox{$\;\stackrel{\textstyle>}{\sim}\;$}}
\def\alp{\bar{\alpha}_s}
\def\w{\omega}
\def\g{\gamma}

\def\bchi{\bar{\chi}}

\def\chilo{\chi_{{\rm veto}}^{{\rm LO}}}
\def\chinlo{\chi_{{\rm veto}}^{{\rm NLO}}}
\def\chiv{\chi_{\rm veto}}

\begin{document}

\begin{flushright}
CERN-TH/99-64 \\
MC-TH-99-04\\
SHEP-99/02\\
March 1999\\
\end{flushright}
\begin{center}
\vspace*{2cm}

%changed the title(dar)
{\Large {\bf Rapidity veto effects in the NLO BFKL equation}} \\

\vspace*{1cm}

J.R.~Forshaw$^1$\footnote{On leave of absence from $^3$}, 
D.A.~Ross$^2$ and A.~Sabio Vera$^3$

\vspace*{0.5cm}
$^1$Theory Division, CERN,\\
1211 Geneva 23, Switzerland.\\
\vspace*{0.2cm}
$^2$Department of Physics,\\
University of Southampton,\\
Southampton, SO17 1BJ. England.\\ 
\vspace*{0.2cm}
$^3$Department of Physics and Astronomy,\\
University of Manchester,\\
Manchester. M13 9PL. England.

\end{center}
\vspace*{5cm}
\begin{abstract}
We examine the effect of suppressing the emission of gluons which are close
by in rapidity in the BFKL framework. We show that, after removing the
unphysical collinear logarithms which typically arise in formally higher 
orders of the perturbative expansion, the effect of the rapidity veto is
greatly reduced. This is an important result, since it supports the use of
multi-Regge and quasi-multi-Regge kinematics which are implemented in the
leading and next-to-leading order BFKL formalism.    
\end{abstract}
\newpage

\section{Eliminating unphysical logarithms}
In the case of fixed coupling the leading eigenvalue of the BFKL kernel, to 
next-to-leading logarithmic accuracy, is determined by solving the equation
\begin{equation}
\w = \chi(\g,\alp) = \alp \chi_0(\g) + \alp^2 \chi_1(\g) \label{NLO}
\end{equation}
where $\alp = N_c \alpha_s / \pi$. The leading order kernel is given by
\begin{equation}
\chi_0(\g) = 2 \psi(1) - \psi(\g) - \psi(1-\g)
\end{equation}
and $\chi_1$ has recently become available \cite{FL,CC}:
\begin{eqnarray}
\chi_1(\gamma) &=& -\frac{1}{8} \left(\frac{11}{3}-\frac{2 n_f}{3 N_c} \right)
\chi_0(\g)^2 + \left( \frac{67}{36} - \frac{\pi^2}{12} - \frac{5 n_f}{18 N_c}
\right) \chi_0(\g) + \frac{3}{2} \zeta(3) \nonumber \\ &-&
\frac{\pi^2 \cos \pi \g}{4 (1-2\g) \sin^2 \pi \g} \left( 3 + \left(
1+\frac{n_f}{N_c^3} \right) \frac{2+3 \g(1-\g)}{(3-2\g)(1+2\g)} \right)
\nonumber \\ &+& \frac{\psi''(\g) + \psi''(1-\g)}{4} + \frac{\pi^3}
{4 \sin \pi \g} - \phi(\g) 
\end{eqnarray}
where
\begin{equation}
\phi(\g) = \sum_{n=0}^{\infty} (-1)^n \left[ \frac{ \psi(n+1+\g)-\psi(1)}
{(n+\g)^2} + \frac{\psi(n+2-\g)-\psi(1)}{(n+1-\g)^2} \right]. 
\end{equation}
We take $n_f=3$ although our results depend only very weakly upon $n_f$.

For our purposes, it is sufficient to consider the gluon Green's 
function:
\begin{equation}
f(s,k_1,k_2) = \int \frac{d\w}{2 \pi i} \frac{d\g}{2 \pi i} \ \left( 
\frac{s}{s_0} \right)^{\w} \left( \frac{k_1^2}{k_2^2} \right)^\g \ 
\frac{1}{\w - \chi(\g,\alp)} \label{Green}
\end{equation}
With $s_0 = k_1 k_2$ the kernel is as given explicitly above.

It has been pointed out that the NLO kernel induces unphysical logarithms in 
the ratio $k_1/k_2$ at NNLO and beyond \cite{CC,Salam}. If $k_1 \gg k_2$ then 
the unphysical logs are induced by the singular behaviour of the kernel at 
$\g = 0$, i.e. $\chi_0 \approx 1/\g$ and $\chi_1 \approx -1/(2 \g^3)$ (the 
symmetry under interchange of $k_1$ and $k_2$ is reflected in the fact that 
the kernel is even about $\g=1/2$). Although these 
logarithms lie formally beyond the boundary of the NLO approach, they can 
induce spurious large effects, especially if the external momenta are strongly
ordered ($k_1 \gg k_2$ or $k_1 \ll k_2$). For consistency with the DGLAP 
approach we know that they must really be absent. At NLO all spurious logs 
are indeed cancelled. Salam demonstrated how to extend the NLO kernel 
so as so guarantee the cancellation of unphysical logs to all orders. 
The prescription for modifying the kernel is ambiguous and Salam investigated 
a variety of different schemes \cite{Salam}. Let us summarise the four
schemes used in \cite{Salam}.

\begin{flushleft}

%changed chi_0+\psi(\g)+\psi(1-\g) to 2\psi(1)
 In scheme 1, one replaces $\chi(\g,\alp)$ of (\ref{NLO}),(\ref{Green}) with
\begin{eqnarray}
\bchi(\g,\w,\alp) &=& \alp [ \ 
2\psi(1) - \psi(\g+\w/2) - \psi(1-\g+\w/2) \ ] \nonumber \\ &+& 
\alp^2 \Big[ \chi_1(\g) + \frac{1}{2} \chi_0(\g)(
\psi'(\g) + \psi'(1-\g)) \nonumber \\ &+& 
\left( \delta_1 + \frac{\pi^2}{6} \right) 
[\psi(\g)+\psi(1-\g) - \psi(\g+\w/2) - \psi(1-\g+\w/2) ] 
\nonumber \\&-& \delta_2 [\psi'(\g) + \psi'(1-\g) - \psi'(\g+\w/2) - 
\psi'(1-\g+\w/2) ] \Big]
\end{eqnarray}
where
\begin{eqnarray}
\delta_1 &=& - \frac{5 n_f}{18 N_c} - \frac{13 n_f}{36 N_c^3} 
\nonumber \\ 
\delta_2 &=& -\frac{11}{8} + \frac{n_f}{12 N_c} - \frac{n_f}{6 N_c^3}
\end{eqnarray} 
are the coefficients of the singular parts of the NLO kernel, i.e.
\begin{equation}
{\rm Lim}_{\ \g \to 0} \ \chi_1(\g) = -\frac{1}{2\g^3} + \frac{\delta_2}{\g^2} 
+ \frac{\delta_1}{\g}.
\end{equation}
This new kernel is identical to the NLO kernel to order $\alp^2$ but is 
completely free from any singularities as $\g \to 0,1$.

The leading $\w$-plane singularity is found by solving 
\begin{eqnarray}
\w = \bchi(\g,\w,\alp).
\end{eqnarray} 

In scheme 2, the new kernel is
\begin{eqnarray}
\bchi(\g,\w,\alp) &=& \alp \left[ \chi_0(\g) - \frac{1}{\g} -\frac{1}{1-\g}
+ \frac{1}{\g+\w/2} + \frac{1}{1-\g+\w/2} \right] \nonumber \\
&+& \alp^2 \left[ \chi_1(\g) + \frac{1}{2}\chi_0(\g) \left(
\frac{1}{\g^2} + \frac{1}{(1-\g)^2} \right) \right. \nonumber \\
&-& \left. \left( \delta_1 + \frac{1}{2} \right) \left(\frac{1}{\g} + 
\frac{1}{1-\g}  - \frac{1}{\g+\w/2} - \frac{1}{1-\g+\w/2} \right) \right.
\nonumber \\ &-& \left. \delta_2 \left( \frac{1}{\g^2} + \frac{1}{(1-\g)^2}  - 
\frac{1}{(\g+\w/2)^2} - \frac{1}{(1-\g+\w/2)^2} \right) \right].
\end{eqnarray} 

In scheme 3, it is
%changed chi_0+\psi(\g)+\psi(1-\g) to 2\psi(1)
\begin{eqnarray}
\bchi(\g,\w,\alp) &=& \alp (1 - \alp A) [
2 \psi(1) -  
\psi(\g+\w/2 + \alp B) - \psi(1-\g+\w/2 + \alp B) ]  \nonumber \\ &+& 
\alp^2 \left[ \chi_1(\g) + \left( \frac{1}{2} \chi_0(\g) + B \right)
(\psi'(\g) + \psi'(1-\g))  + A \chi_0(\g) \right] 
\end{eqnarray}
where 
\begin{eqnarray}
A &=& -\delta_1 - \frac{\pi^2}{6} \nonumber \\
B &=& -\delta_2
\end{eqnarray}

Finally, in scheme 4, the new kernel is
\begin{eqnarray}
\bchi(\g,\w,\alp) &=& \alp \left[ \chi_0(\g) - \frac{1}{\g} -\frac{1}{1-\g}
\right. \nonumber \\ &+& \left. (1 - \alp A')\left( 
\frac{1}{\g+\w/2+\alp B} +  \frac{1}{1-\g+\w/2+\alp B}  \right) \right] 
\nonumber \\
&+& \alp^2 \left[ \chi_1(\g) + \left(B+\frac{1}{2}\chi_0(\g)\right) \left(
\frac{1}{\g^2} + \frac{1}{(1-\g)^2} \right) + A' \left(\frac{1}{\g} + 
\frac{1}{1-\g} \right) \right] 
\end{eqnarray} 
where $B$ is as above and
where 
\begin{equation}
A' = -\delta_1 - \frac{1}{2}.
\end{equation}

\end{flushleft}

\begin{figure} 
\centerline{\epsfig{file=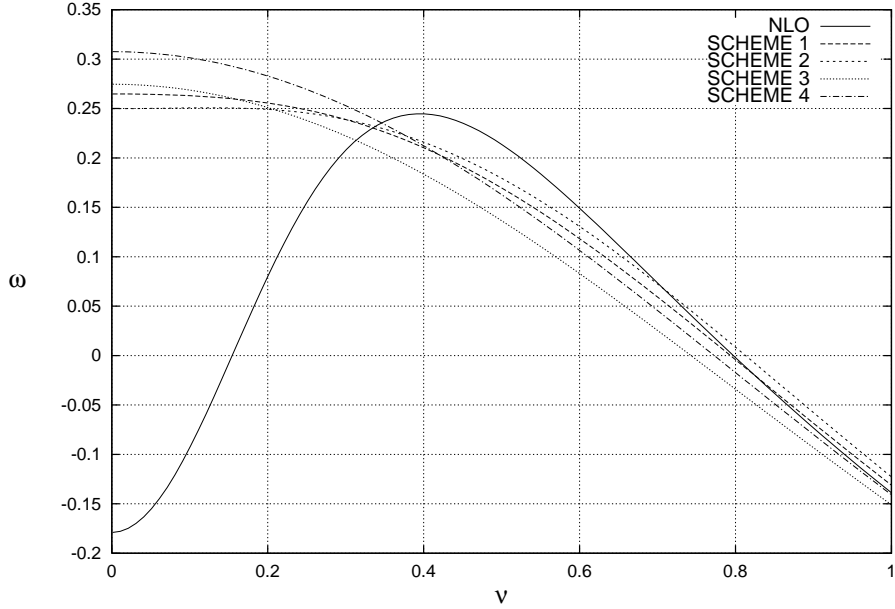,height=12cm,angle=-90}}
\caption{Behaviour of the collinear improved kernels described in the text}
\label{Salfig}
\end{figure}

We note that other schemes have been advocated \cite{Colf}.
It was demonstrated in \cite{Salam} that the all orders kernel is much 
better behaved than the NLO kernel. In particular, in schemes 3 and 4 it was 
shown that the subsequent integral over $\gamma$ is again dominated by the 
saddle point at $\gamma = 1/2$ and that the leading eigenvalue of the kernel 
does not pick up huge higher order corrections (although they do still lead 
to a significant reduction in its value). To illustrate these points, 
in Fig.~\ref{Salfig} we show the behaviour of the all 
order kernels as a function of $\nu$
 where $\gamma = 1/2 + i \nu$, with $\alp=0.2$, 
and compare with the behaviour of the NLO kernel (which no longer
has the saddle point at $\gamma=1/2$ \cite{DAR}).
It can be seen that for these modest values of $\alp$ the result
is not very sensitive to the scheme used.  In this paper we
follow Salam in maintaining that schemes 3 and 4 are the most realistic 
schemes and focus upon them.

\section{The rapidity veto}
Imposing the constraint that subsequent gluon emissions must be separated by 
some minimum interval in rapidity, $b$, upon the LO BFKL equation leads to
the new kernel:
\begin{equation}
\chilo(\g,\w,\alp) = \alp e^{-b \w} \chi_0(\g). \label{LOveto}
\end{equation}
This kernel encodes the rapidity veto to all orders in $\alp$.

It has been suggested that a veto of $b \sim 1$ may well be a good
approximation to real physics and as such may account for a large part of
the NLO corrections to the leading eigenvalue of the kernel 
\cite{Lund,Lund1,Lip}.
The effect of imposing the veto on the LO BFKL equation leads to a leading 
eigenvalue which is determined by the solution to
\begin{equation}
\w = \alp e^{-b \w} \chi_0(\g). \label{LOveto1}
\end{equation}
Expanding in $\alp$ gives
\begin{eqnarray}
\w &\approx& (1 - b~\w) \ \w_0 \nonumber \\
\w &\approx& \frac{\w_0}{1+b~\w_0} \nonumber \\
   &\approx& \w_0 (1- b~\w_0)
\end{eqnarray}
and $\w_0 = \alp \chi_0(1/2)$, which is relevant if we assume that the 
saddle point at $\g=1/2$ is reliable. This is to be compared with the
NLO kernel, which gives
\begin{equation} 
\w = \w_0 (1 - 2.4~\w_0).
\end{equation}
If a veto of 2 units were physics then we see that it saturates most of the
NLO correction, leaving behind a genuinely small correction. The study of
a rapidity veto has been pursued in \cite{Schmidt}. 

At NLO, the imposition of the veto leads to
\begin{equation}
\chinlo(\g,\w,\alp) = \alp e^{-b \w} [ \chi_0(\g) + \alp \chi_1(\g) +
\alp b \chi_0(\g)^2 ].
\end{equation}

\begin{figure} 
\centerline{\epsfig{file=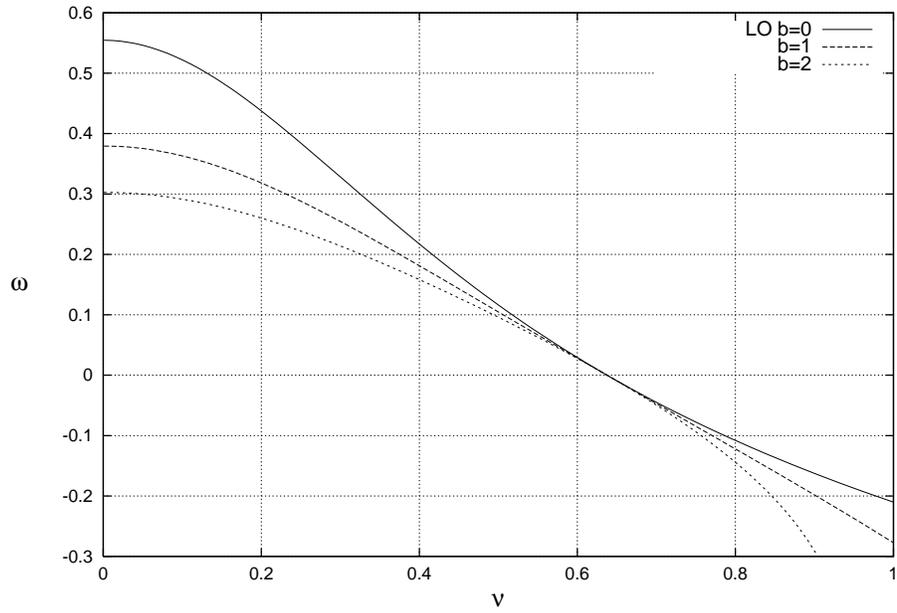,height=12cm,angle=-90}}
\caption{Dependence of the LO kernel plus veto upon $\nu$, 
for $\alp=0.2$} %added value of alpha (dar)
\label{vetofiga}
\end{figure}

\begin{figure} 
\centerline{\epsfig{file=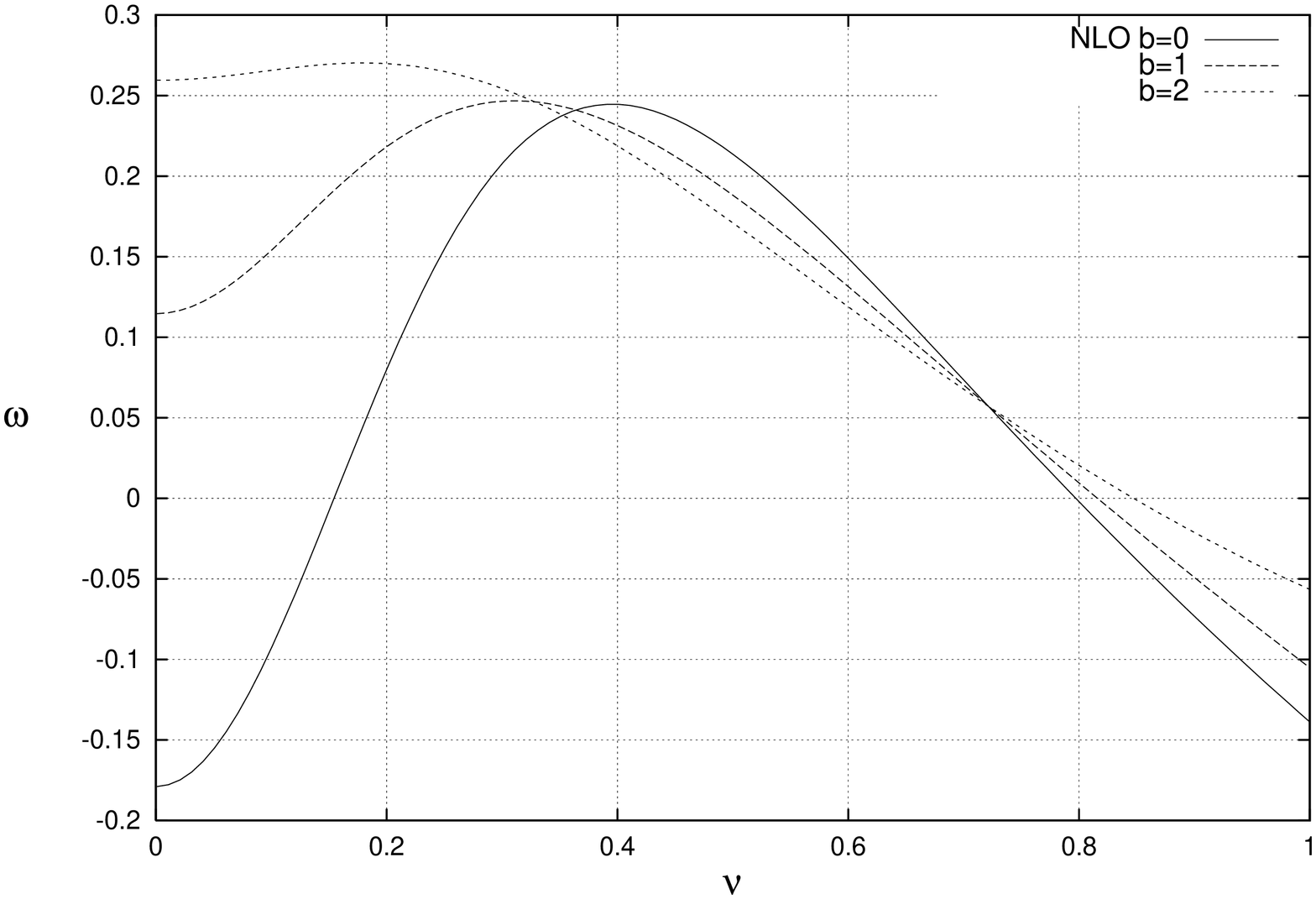,height=12cm,angle=-90}}
\caption{Dependence of the NLO kernel plus veto upon $\nu$,
for $\alp=0.2$} %added value of alpha (dar)
\label{vetofigb}
\end{figure}

\begin{figure} 
\centerline{\epsfig{file=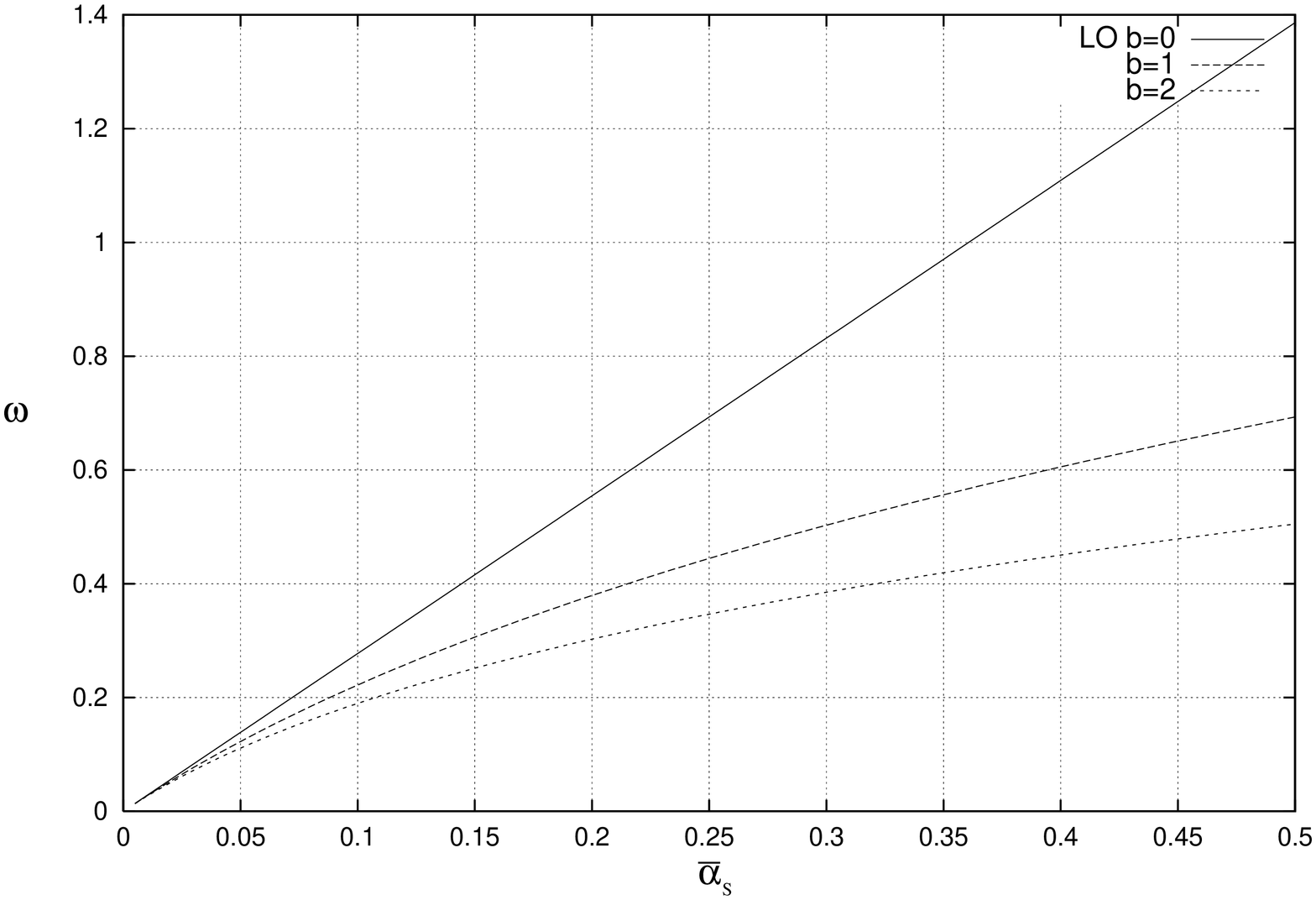,height=12cm,angle=-90}}
\caption{Dependence of the LO kernel plus veto upon $\alp$,
for $\nu=0$} %added value of nu (dar)
\label{vetofigc}
\end{figure}

\begin{figure} 
\centerline{\epsfig{file=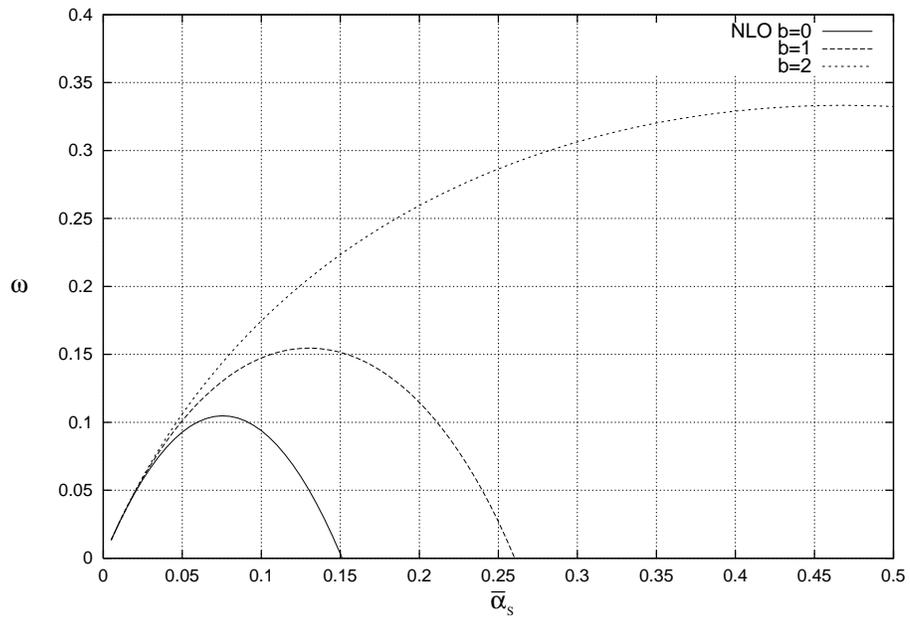,height=12cm,angle=-90}}
\caption{Dependence of the NLO kernel plus veto upon $\alp$,
for $\nu=0$} %added value of nu (dar)
\label{vetofigd}
\end{figure}

In Figs.~\ref{vetofiga}--\ref{vetofigd} we show the effect of imposing the 
veto on the LO and NLO  BFKL equations. We show seperately the dependence of 
the kernels upon $\nu$ and the variation of the kernels evaluated at $\nu=0$ 
with $\alp$.  In each case we show results for three different 
values of $b$, $b=0$ (no veto), $b=1$ and $b=2$. 
Clearly the imposition of the veto has a dramatic
effect upon the behaviour of the kernel.
 
In this paper, we wish to investigate the effect of imposing a rapidity veto
in conjunction with the removal of the unphysical logarithms. Our philosophy 
is as follows. If, after removing the unphysical logarithms, the imposition 
of a rapidity veto has little effect then we have internal consistency. 
This is because the BFKL approach relies on the assumption of multi-Regge
kinematics, i.e. that successive emissions are infinitely far apart in
rapidity, at LO and it assumes that subsequent corrections should be small.
It follows that cutting out emissions which are nearby in rapidity should
have a relatively small effect. We show that this is indeed the case.

The rapidity veto can be implemented in a way which ensures that the 
cancellation of the unphysical logarithms is not disturbed. The procedure
is scheme dependent, just as in the case without any veto. 

\begin{flushleft}

For scheme 3 we have
\begin{eqnarray}
\chiv(\g,\w,\alp) &=& \alp e^{-b \w} \big\{ (1 - \alp A) [ 
2\psi(1) \nonumber \\ &-& \psi(\g+\w/2 + \alp B') - 
\psi(1-\g+\w/2 + \alp B') ] \nonumber \\ & &
\hspace*{-2cm} +   \alp \left[ \chi_1(\g) + b \chi_0(\g)^2 + 
\left( \frac{1}{2} \chi_0(\g) + B' \right)
(\psi'(\g) + \psi'(1-\g))  + A \chi_0(\g) \right] \big\} 
\end{eqnarray}
where $A$ is as before but $B'$ is now
\begin{equation}
B' = -\delta_2 - b, \end{equation}
which is required to ensure that the introduction of the rapidity
gap does not destroy the elimination of the unphysical
logarithms. %JRF -- removed italics

For scheme 4 we have:
\begin{eqnarray}
\chiv(\g,\w,\alp) &=& \alp e^{-b \w} \left\{ \left[ \chi_0(\g) - 
\frac{1}{\g} -\frac{1}{1-\g} \right. \right. \nonumber \\ &+& \left. \left. 
(1 - \alp A')\left( \frac{1}{\g+\w/2+\alp B'} +  
\frac{1}{1-\g+\w/2+\alp B'}  \right) \right] \right. \nonumber \\ 
& & \hspace*{-3cm} + 
\left. \alp \left[ \chi_1(\g) + b \chi_0(\g)^2 + 
\left(B'+\frac{1}{2}\chi_0(\g)\right) \left(
\frac{1}{\g^2} + \frac{1}{(1-\g)^2} \right) + A' \left(\frac{1}{\g} + 
\frac{1}{1-\g} \right) \right] \right\}. 
\end{eqnarray} 

These new kernels implement the veto, are matched to NLO and free from
any singularities as $\g \to 0,1$.

\end{flushleft}

\begin{figure} 
\centerline{\epsfig{file=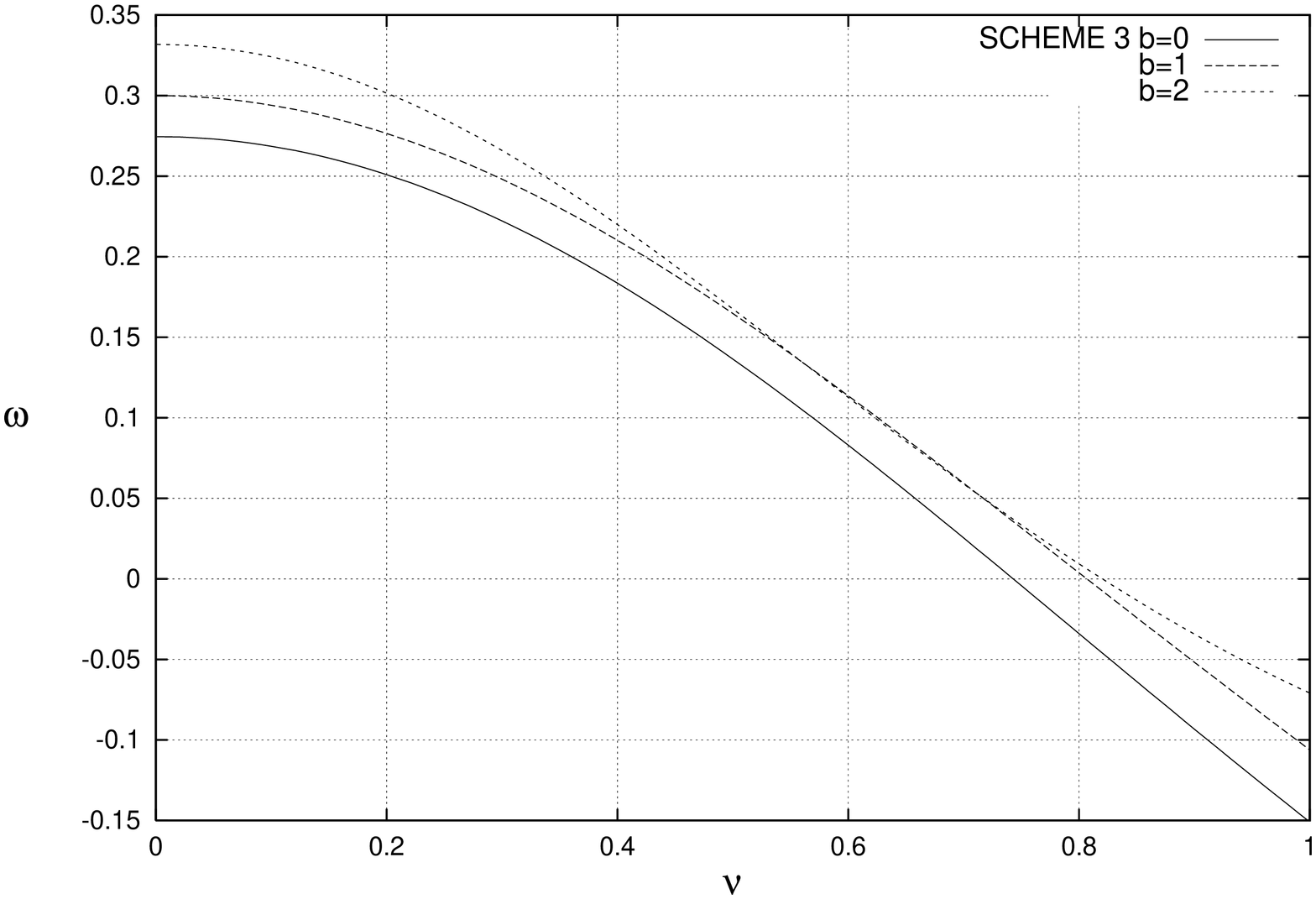,height=12cm,angle=-90}}
\caption{Dependence of the `Scheme 3 plus veto' NLO kernel upon $\nu$,
for $\alp=0.2$} %added the value of alpha (dar)
\label{s3figa}
\end{figure}

\begin{figure} 
\centerline{\epsfig{file=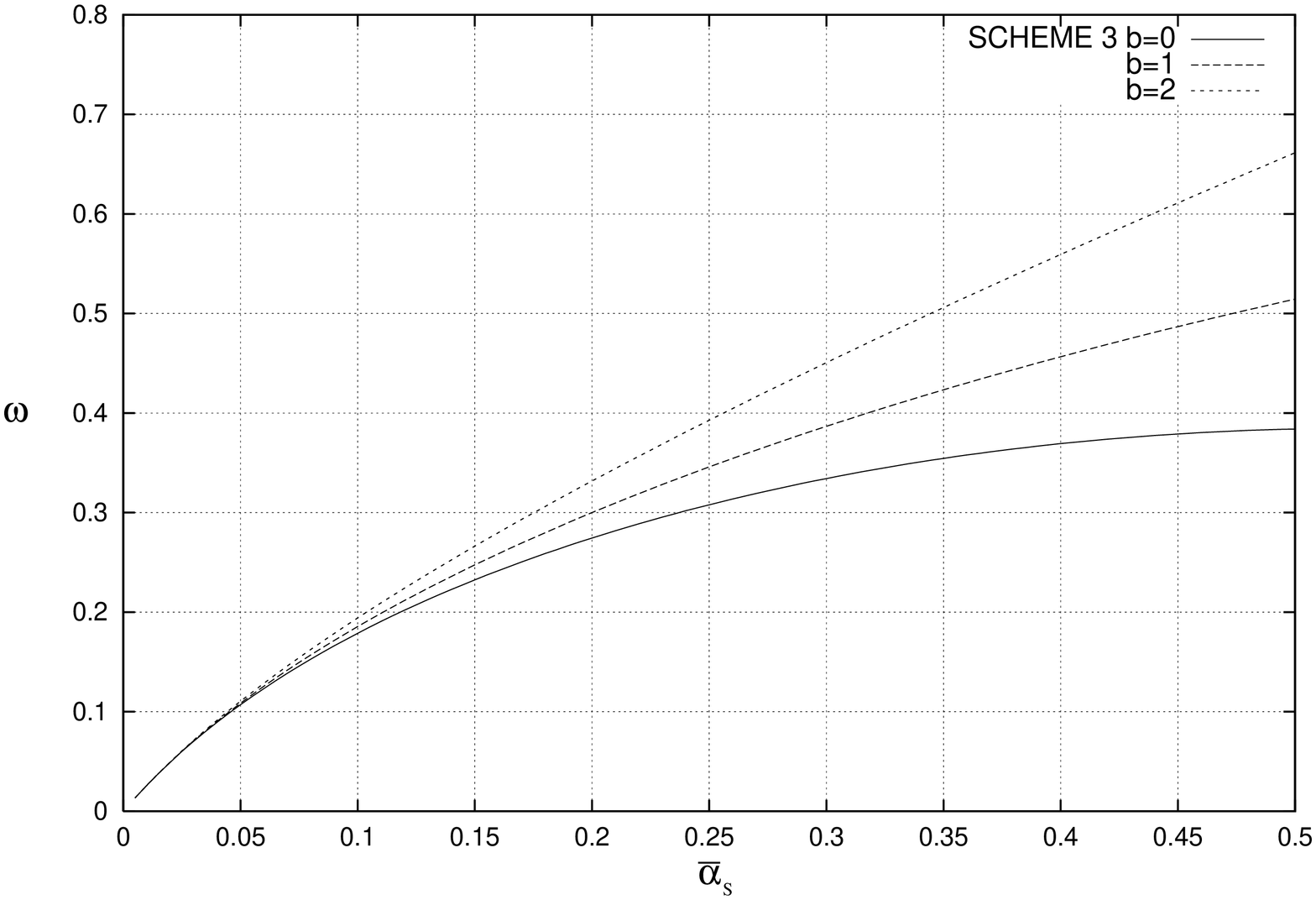,height=12cm,angle=-90}}
\caption{Dependence of the `Scheme 3 plus veto' NLO kernel upon $\alp$,
for $\nu=0$} %\added the value of nu (dar)
\label{s3figb}
\end{figure}

\begin{figure} 
\centerline{\epsfig{file=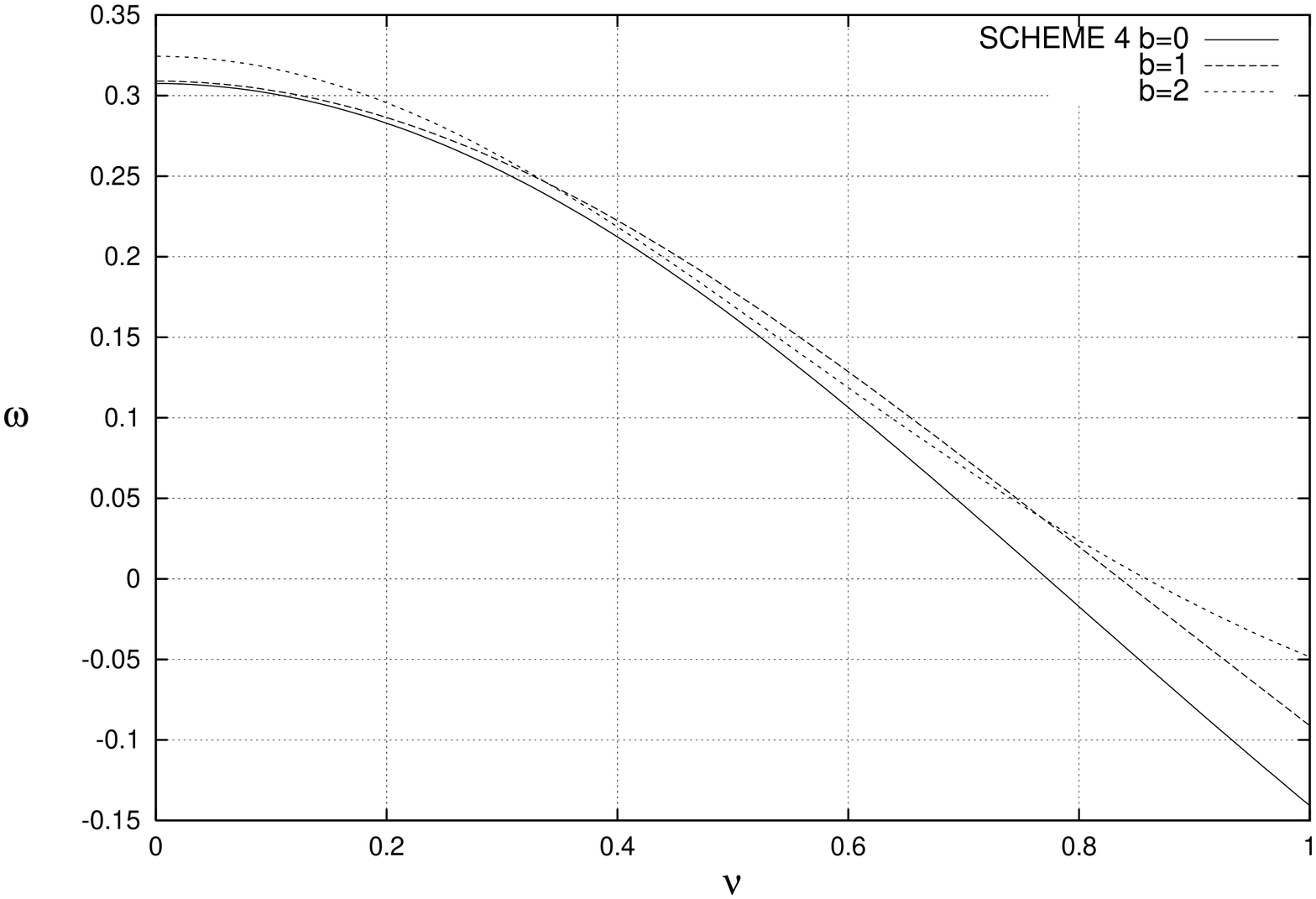,height=12cm,angle=-90}}
\caption{Dependence of the `Scheme 4 plus veto' NLO kernel upon $\nu$,
for $\alp=0.2$} %added the value of alpha (dar)
\label{s4figa}
\end{figure}

\begin{figure} 
\centerline{\epsfig{file=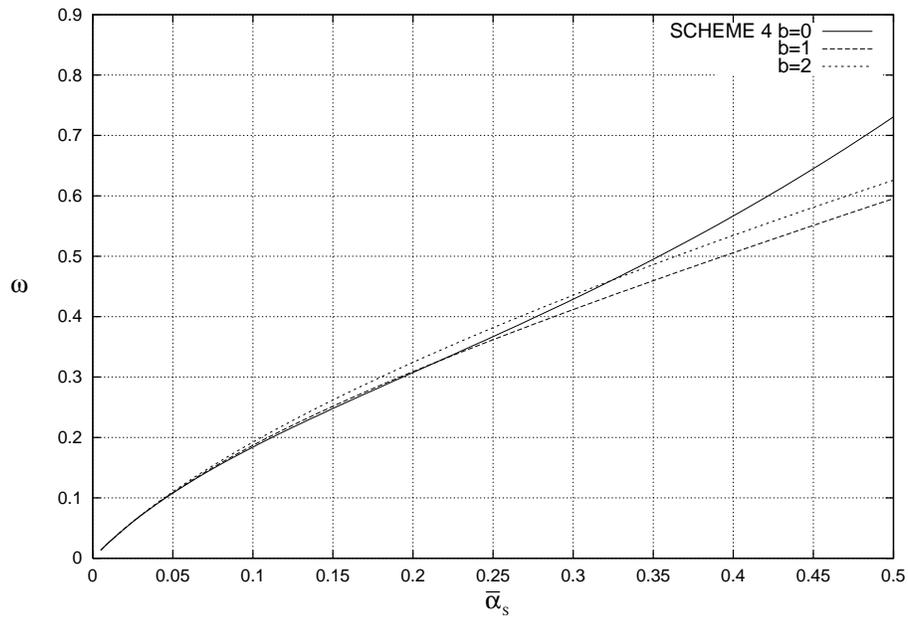,height=12cm,angle=-90}}
\caption{Dependence of the `Scheme 4 plus veto' NLO kernel upon $\alp$,
for $\nu=0$} 
\label{s4figb}
\end{figure}

In Figs.~\ref{s3figa},\ref{s3figb}, we do as in 
Figs.~\ref{vetofiga}--\ref{vetofigd} but using the new Scheme 3 kernel and 
in Fig.~\ref{s4figa},\ref{s4figb} we show the results for the Scheme 4 
kernel. Having subtracted the
unphysical logarithms we can see clearly that the dependence upon the 
rapidity veto is greatly diminished. We interpret this as supporting the
use of (quasi-)multi-regge kinematics. 

In the case of the pure LO and NLO kernels and their collinear corrected
variants, the behaviour of the kernel near $\nu=0$ drives the asymptotic
($s/s_0 \to \infty$) behaviour of the Green function of (\ref{Green}). 
However, the situation is far from clear after implementing the veto.  

For definiteness, let us take the kernel of (\ref{LOveto}). We note that
the solution to (\ref{LOveto1}) develops cuts along the line $\gamma=1/2$
with branch points at $\pm \nu_0$: 
\begin{equation}
\chi_0 \left( \frac{1}{2} + i \nu_0 \right) =  -\frac{e^{-1}}{\alp b }.
\end{equation}
However, this cut has no physical origin. It arises 
because we summed over arbitrarily large numbers of gluon emissions with
each emission some minimum distance in rapidity from its neighbours. For
any finite energy this is not possible. The result of truncating the number
of emissions is to remove the cut. To see this, we first note that after 
imposing the veto and performing the integral over $\w$ the Green function of 
(\ref{Green}) becomes
\begin{equation}
f(s,k_1,k_2) = \int \frac{d\g}{2 \pi i} \left( \frac{s}{s_0} \right)^{\w_0(\g)}
\left( \frac{k_1^2}{k_2^2} \right)^{\g} \frac{1}{1 + b \w_0(\g)}
\label{exact}
\end{equation}
where $\w_0(\g)$ is the solution to (\ref{LOveto1}). We can now expand the
RHS as a power series in $\alp$ \cite{Schmidt}:
\begin{equation}
\frac{e^{\w_0(\g) y}}{1 + b \ \w_0(\g)}
= \sum_{n=0}^{\infty} \frac{ [\alp \chi_0(\g) \ (y - n b)]^n }{n!}
\label{series}
\end{equation}
where $y \equiv \ln (s/s_0)$. Limiting the number of emitted gluons forces
us to truncate the summation of the RHS of (\ref{series}) and since 
the only singularities of $\chi_0(\g)$ are poles at $\g = 0,-1,-2,...$ it
follows that the cut is absent. 

\begin{figure} 
\centerline{\epsfig{file=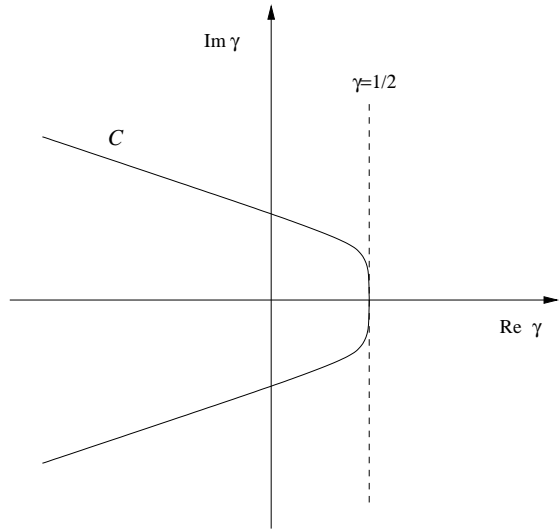,height=7cm,angle=0}}
\caption{A suitable contour of integration for $k_1 > k_2$.}
\label{contour}
\end{figure}

Our claim that imposing the veto has a small effect is still not complete.
We must also show that the asymptotic behaviour is driven by the behaviour of
the kernel near $\nu = 0$. This is not as obvious as in the cases without a
veto since the solution to (\ref{LOveto1}), $\w_0(\g)$, does not fall
monotonically as $\g \to \infty$. In fact it grows slowly as 
$$ \w_0(\g) \sim \ln [\ln \g(1-\g)]. $$ 
However, the presence of the $(k_1^2/k_2^2)^\g$ term in the integral 
ensures that the integrand falls monotonically from its maximum in the
vicinity of $\nu=0$. For example, for $k_1 > k_2$ any $\g$-plane contour 
which heads to Re~$\g \to - \infty$, e.g. the contour $C$ shown in 
Fig.~\ref{contour}, is suitable. For large
enough $y$ we can therefore approximate (\ref{exact}) by
\begin{equation}
f(s,k_1,k_2) \approx \frac{e^{\tilde{\w}_0 y}}{1+b \ \tilde{\w}_0}
\int_{-\infty}^{\infty}  \frac{d\nu}{2 \pi} 
\left( \frac{k_1^2}{k_2^2} \right)^{1/2+i \nu} 
e^{-A y \nu^2} \label{dar3}
\end{equation}
where $\omega_0(\g) \approx \tilde{\w}_0 - A \nu^2$ in the vicinity of
$\nu = 0$.
The argument runs through in precisely the same fashion for the other 
kernels which include a veto. We have demonstrated numerically
that the results obtained from the approximation (\ref{dar3}) 
are compatible with the results obtained from truncating the series
(\ref{series}) and inserting into (\ref{exact}).

Thus, for large enough $s/s_0$ the truncated sum can be approximated by
the infinite sum\footnote{In practice, the infinite sum is a good 
approximation down to quite low values of $y$, e.g. $y \gapproxeq 1$ for
$\alp = 0.2$.}
and the integral over $\g$ can be approximated using
the saddle point method, i.e. the asymptotic behaviour is driven by the
value of the kernel at $\nu=0$ (i.e. $\g = 1/2)$. 
It is this precisely this region of the kernel which we have shown to be 
unaffected by the imposition of the veto.

\section{Conclusions}
We have demonstrated that, after taking care to eliminate unphysical collinear
logarithms in the BFKL formalism, the resulting kernel is insensitive
to the imposition of the restriction that successive gluon emissions not
be too close together in rapidity. Insensitivity to such a veto supports the
application of the (quasi-)multi-Regge kinematics provided that the 
logarithms in transverse momenta are treated in a consistent manner.

\section{Acknowledgements}
We thank Lev Lipatov and Gavin Salam for helpful discussion.
ASV would like to thank PPARC for the award of a Studentship.
This work was supported in part by the EU Fourth Programme `Training and 
Mobility of Researchers', Network `Quantum Chromodynamics and the Deep 
Structure of Elementary Particles', contract FMRX-CT98-0194 (DG 12-MIHT).

\end{document}